\documentclass[aps,prl,reprint,superscriptaddress,showpacs,floatfix]{revtex4-1}

\usepackage[pdftex]{graphicx}
\usepackage{amsmath}

\begin{document}

\title{Quantum confined electronic states in atomically well-defined graphene nanostructures}
\author{Sampsa K. H\"{a}m\"{a}l\"{a}inen }
\affiliation{Department of Applied Physics, Aalto University School of Science, 00076 Aalto, Finland}
\author{Zhixiang Sun}
\author{Mark P. Boneschanscher}
\affiliation{Condensed Matter and Interfaces, Debye Institute for Nanomaterials Science,
Utrecht University, PO Box 80000, 3508 TA Utrecht, the Netherlands}
\author{Andreas Uppstu}
\affiliation{Department of Applied Physics, Aalto University School of Science, 00076 Aalto, Finland}
\author{Mari Ij\"as}
\author{Ari Harju}
\affiliation{Department of Applied Physics, Aalto University School of Science, 00076 Aalto, Finland}
\affiliation{Helsinki Institute of Physics, Aalto University School of Science, 00076 Aalto, Finland}
\author{Dani\"{e}l Vanmaekelbergh}
\affiliation{Condensed Matter and Interfaces, Debye Institute for Nanomaterials Science,
Utrecht University, PO Box 80000, 3508 TA Utrecht, the Netherlands}
\author{Peter Liljeroth}
\email[]{peter.liljeroth@aalto.fi}
\affiliation{Department of Applied Physics, Aalto University School of Science, 00076 Aalto, Finland}
\affiliation{Condensed Matter and Interfaces, Debye Institute for Nanomaterials Science,
Utrecht University, PO Box 80000, 3508 TA Utrecht, the Netherlands}
\affiliation{Low Temperature Laboratory, Aalto University School of Science, PO Box 15100, 00076 Aalto, Finland}

\date{\today}

\begin{abstract}
Despite the enormous interest in the properties of graphene and the potential of graphene nanostructures in electronic applications, the study of quantum confined states in atomically well-defined graphene nanostructures remains an experimental challenge. Here, we study graphene quantum dots (GQDs) with well-defined edges in the zigzag direction, grown by chemical vapor deposition (CVD) on an iridium(111) substrate, by low-temperature scanning tunneling microscopy (STM) and spectroscopy (STS). We measure the atomic structure and local density of states (LDOS) of individual GQDs as a function of their size and shape in the range from a couple of nanometers up to ca. 20 nm. The results can be quantitatively modeled by a relativistic wave equation and atomistic tight-binding calculations. The observed states are analogous to the solutions of the text book "particle-in-a-box" problem applied to relativistic massless fermions.
\end{abstract}

\pacs{73.21.La 73.22.Pr 73.63.Kv 81.05.ue}

\maketitle

Graphene, a monolayer of carbon atoms that is a 2-D metal where the charge carriers behave as massless relativistic electrons has attracted enormous scientific and technological interest. \cite{Geim2007,CastroNeto2009,Li2009,Schwierz2010}. Despite the potential of graphene nanostructures in electronic applications \cite{Han2007,Rycerz2007,Ponomarenko2008,Jiao2009,Kosynkin2009,Ritter2009,Cai2010,Sprinkle2010}, the study of quantum confined electronic states in atomically well-defined graphene nanostructures remains an experimental challenge. Basic questions, such as the relation between the atomic configuration of graphene nanostructures and the spatial distribution and energy of their electronic states have not been experimentally addressed.

In previous experiments, macroscopic graphene sheets have been studied by scanning tunneling microscopy (STM) and spectroscopy (STS), focusing on the electronic structure and scattering processes in epitaxial graphene \cite{Rutter2007,Sun2011} and the density of states and charge puddles in graphene sheets deposited on an insulator \cite{Martin2008,Zhang2008,Deshpande2009,Zhang2009}. It is, however, also possible to grow much smaller graphene nanostructures (graphene quantum dots, GQDs) by chemical vapor deposition (CVD), and characterize them with scanning probe methods \cite{Ritter2009,Wang2011,Eom2009,Coraux2009}.

In this report, we present low-temperature STM and STS experiments on GQDs with well-defined atomic structures grown by CVD on an Ir(111) substrate. We can readily access individual GQDs and measure their atomic structure with STM. Measurement of the local density of states (LDOS, proportional to the $dI/dV_\mathrm{b}$ signal) allows us to probe the spatial structure and energy of the quantum confined energy levels for GQDs with variable size and shape. The measured LDOS maps can be reproduced by tight-binding (TB) calculations, where we use the exact atomic structure of the GQDs as determined by STM, and by the Klein-Gordon (KG) equation, which is a continuum model describing particles with linear dispersion.

\begin{figure}
\includegraphics{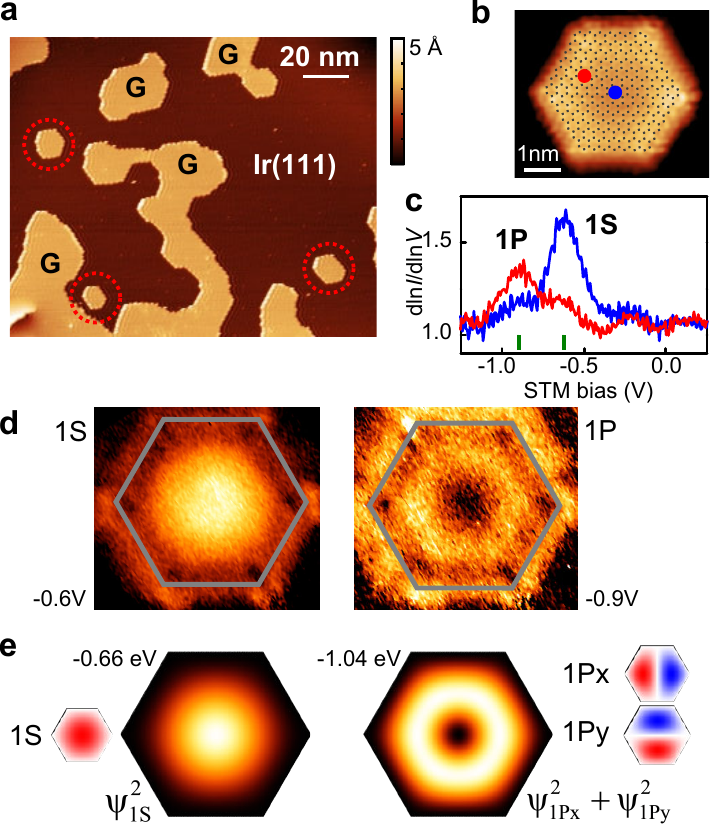}
\caption{(Color online) STM imaging and spectroscopy on GQDs grown by CVD on Ir(111). (a) Large scale STM image of graphene islands (G) on an iridium(111) substrate (acquired at $I$ = 40 pA and $V_\mathrm{b}$ = 1.0 V). Small graphene QDs have been indicated by red circles. (b) STM topography of a small GQD (-0.05 V / 100 pA) with an overlaid atomic model which has perfect hexagonal symmetry with 7 benzene ring long edges. (c) dln$I$/dln$V_\mathrm{b}$ spectra measured on the points indicated in (b), the green lines indicate the bias voltages corresponding to the LDOS maps shown in panel (d). (d) Measured LDOS maps (gray line denotes the edges of the GQD) at bias voltages corresponding to the two resonances in the spectrum shown in panel (c). (e) The corresponding LDOS maps calculated for a particle in a box at the indicated energies and the underlying eigenstates.}\label{Fig1}
\end{figure}

The Ir(111) surface was cleaned by sputtering at 1100 K and annealing at 1500 K. After the sample had cooled below 570 K, ethylene was deposited ($3\times10^{-6}$ mbar for 10 s) on the surface.  The GQD size could then be controlled by the growth temperature \cite{Coraux2009}: larger (smaller) GQDs were grown by heating the sample to 1470 K (1170 K) for 10 s. After the CVD growth of the GQDs, the sample was inserted into a low-temperature STM ($T$ = 4.8 K, Omicron LT-STM) housed within the same ultra-high vacuum system (base pressure $<10^{-10}$ mbar). We used cut PtIr tips and the bias voltage ($V_\mathrm{b}$) was defined as sample voltage with respect to the tip. The $dI/dV_\mathrm{b}$ signal was recorded with a lock-in amplifier by applying a small sinusoidal variation to the bias voltage (typically 30 mV rms at 660 Hz). This gives an energy resolution of ca. 75 meV in our experiments \cite{Morgenstern2003}. To ensure that the modulation would not couple to the feedback loop, a 300 Hz low-pass filter was used for the feedback input. The experimental $dI/dV_\mathrm{b}$ images are averages of the trace and retrace scans (Fig. 1) and of the trace and retrace scans of two consecutive images (up and down, Fig. 2) to increase the signal to noise ratio.

Fig. 1(a) shows a large-scale overview scan of a typical sample. We find interconnected graphene patches (indicated by G) as well as small isolated GQDs (red dotted circles). The CVD growth yields a relatively broad distribution of different GQD sizes ranging from a couple of nanometers up to ca. 20 nm, most of them with a roughly hexagonal shape. All the GQDs have edges in the zigzag direction (corresponding to the close-packed atomic rows of the underlying Ir(111) surface) with a very small roughness. We see kinks of one or two atomic rows at the GQD edges \cite{EPAPS}. Closer inspection of small GQDs at a bias voltage close to zero bias shows that the edges are bright both in the actual STM topography as well as in the simultaneously recorded $dI/dV_\mathrm{b}$ images. These edge states are expected for zigzag edges in graphene \cite{Son2006,Brey2006,Koskinen2008,CastroNeto2009}. More information can be found in the Supplementary Information \cite{EPAPS}.

We now focus on the delocalized, quantum confined states inside the GQDs. We can map the atomic structure of the GQD by STM as shown for a small GQD with perfect hexagonal symmetry with 7 benzene ring long edges in Fig. 1(b). The LDOS can be accessed through $d\ln I/d\ln V_\mathrm{b}$ measurements as shown in Fig. 1(c); we clearly observe an increased and spatially dependent LDOS on the GQDs. There is a pronounced maximum of the LDOS measured in the center of the GQD (blue line in Fig. 1(c)) at a bias of -0.6 V. Moving away from the center of the GQD, the intensity of this peak is reduced, and another resonance emerges at a bias of -0.9 V (red line). We can map the spatial shape of the orbitals responsible for these resonances by measuring the $dI/dV_\mathrm{b}$ signal during STM imaging under constant-current feedback at biases corresponding to the resonances [Fig. 1(d)]. These states have the familiar appearance of the lowest energy levels of the text book particle-in-a-box problem and can be characterized using symmetry labels borrowed from atomic physics. The lowest state has $1S$ symmetry (no nodal planes) and the first excited state is composed of two $1P$ type orbitals ($1P_x$ and $1P_y$) which are degenerate in this case of a perfect hexagonal GQD. STM probes the sum of the squared wavefunctions $\psi_\mathrm{1P_x}^2+\psi_\mathrm{1P_y}^2$ leading to a doughnut shaped $dI/dV_\mathrm{b}$ signal as we observe in the experiment. Comparison of these states with TB calculations can be found in the Supplementary Information \cite{EPAPS}.

We note here that at positive bias, electronic resonances with clear peaks in the $dI/dV_\mathrm{b}$ spectrum cannot be observed \cite{EPAPS}. Based on DFT calculations on Ir(111), there is a dense set of energy bands above the Fermi energy at the K point of Brillouin zone. It is likely that interaction with these states masks the intrinsic graphene states at positive bias \cite{Pletikosic2010}.

These experiments can be reproduced by both TB calculations and by a continuum model for particles with linear dispersion confined to a GQD \cite{EPAPS}. Here we use the KG equation \cite{Heiskanen2008,Barbier2008}
\begin{equation}
-v_\mathrm{F}^2\hbar^2\nabla^2\psi_i=E_i^2\psi_i
\end{equation}
where $v_\mathrm{F}$ is the Fermi velocity ($10^6$ m/s in isolated graphene) and the boundary condition is given by $\psi_i$ =  0 at the edges of the GQD. A more accurate boundary condition would be needed to take into account the sublattice pseudospin and the interaction with the Ir substrate. It is clear that the KG equation cannot be used to model the edge states (in contrast to the Dirac equation and TB calculations \cite{Brey2006,Son2006,CastroNeto2009}). However, as shown below, the LDOS plots from the KG equation are remarkably similar to the TB calculations and the experimental results, although the number of states in a given energy interval is too small. We use the experimentally determined geometries of the GQD in our calculations \cite{EPAPS}. The lowest energy solutions of Eq. (1) are plotted in Fig. 1(e) as the squared wavefunctions corresponding to the experimentally measured $\mathrm{LDOS}\propto\sum_{\delta E}\psi_i^2$, where $\delta E$ is the energy resolution of the experiment \cite{Tersoff1985}.

\begin{figure}
\includegraphics{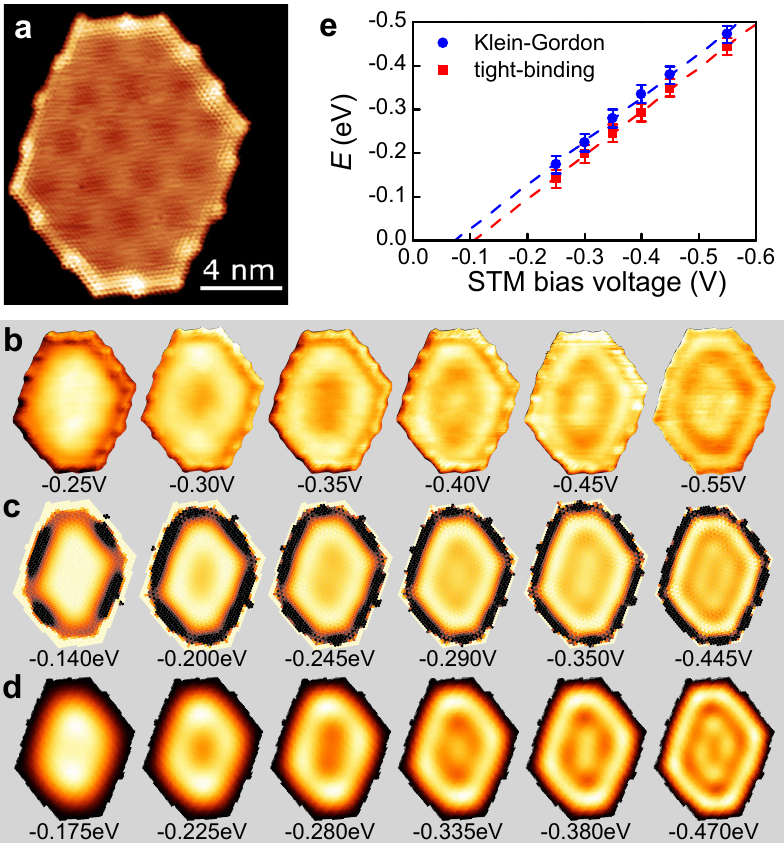}
\caption{(Color online) Detailed comparison between STM and STS experiments and computational results on a large GQD. (a) Atomically resolved STM image of the GQD ($I$ = 3 nA, $V_\mathrm{b}$ = 1 mV). (b) d$I$/d$V_\mathrm{b}$ maps recorded under constant-current STM feedback at the bias voltages indicated in the figure ($I$ = 1 nA). (c,d) Corresponding LDOS plots at the indicated energies calculated using a TB model (c) and the KG equation (d) as described in the text. (e) Correspondence between the experimental and the calculated energies based on TB (red squares) and the KG equation (blue circles) calculated with $v_\mathrm{F}=6.2\times 10^5$ m/s.}\label{Fig2}
\end{figure}

We have measured the LDOS at different bias voltages on a larger GQD shown in Fig. 2(a). The periodic variation with a period of 2.5 nm seen on the topographic STM images is a moir\'e pattern resulting from the lattice mismatch between graphene and Ir \cite{Coraux2009,Sun2011}. The STM contrast results mostly from a small (ca. 30 pm) geometric modulation of the graphene structure \cite{Sun2011}. Our calculations neglecting this moir\'e-induced potential modulation yield quantitative agreement with the experiment and the expected potential modulation due to the moir\'e pattern is small compared to the confinement energy in our GQDs. It has been reported that the size and shape of the GQDs is influenced by the moir\'e pattern and the edges prefer to run along the fcc and hcp regions of the moir\'e \cite{NDiaye2008,Coraux2009}. We also observe GQDs that are smaller than the moir\'e period ($6\times6$ and $7\times7$). For larger GQDs, the kinks on the edges are spaced by roughly one moir\'e period.

The asymmetry of the GQD breaks the degeneracies (e.g. $1P_x$ and $1P_y$ states) of the purely hexagonal GQD. This can be seen in the measured LDOS maps shown in Fig. 2(b) (The Ir substrate has been removed in the images using the simultaneously acquired STM topography image as a mask, images with the background can be found in the Supplementary Information \cite{EPAPS}): after the $1S$ state (bias -0.25 V), we observe increased intensity at the top and bottom end of the GQD consistent with the $1P_y$ envelope wavefunction along the long GQD axis (at -0.30 V). At more negative bias, the $1P_x$ state also contributes and the long GQD edges are brighter (-0.35 V). Subsequently, the next eigenstate becomes relevant, which is seen as an increased intensity in the middle of the QD (bias -0.4 V).

In order to compare experiment and theory in detail, we have generated a series of theoretical LDOS maps, which are calculated as a weighted and broadened sum of squares of TB molecular orbitals (MOs) or KG eigenstates close to a given energy [see Figures 2(c,d)] \cite{EPAPS}. This broadening is justified due to the intrinsic resolution of the measurement (75meV) and the life-time broadening of the states. In the case of the calculations based on the KG equation,the eigenfunctions are given by the solution of Eq. (1) using the overall shape of the GQD. In the TB calculations (we use third-nearest-neighbor TB) \cite{Son2006,CastroNeto2009,Hancock2010}, they correspond to the calculated MOs for the GQD with an exact atomic structure as obtained from experiment [Fig. 2(a)] \cite{EPAPS}. It can be seen that the eigenstates of the KG equation (overall geometry) match with clusters of TB MOs (exact atomic lattice). Furthermore, there is a remarkable agreement in how both calculated LDOS maps evolve with energy and how the experimental conductance maps evolve with the bias.

Based on a comparison between the experimental and computed LDOS maps, we have identified energy / bias voltage pairs that give the same spatial features in the LDOS with an associated error estimate indicated by error bars in Fig. 2(e) \cite{EPAPS}. It is clear that with the Fermi-velocity $v_\mathrm{F}$ as the only adjustable parameter (in the case of TB calculations, $v_\mathrm{F}$ is directly related to the value of the hopping integrals), both calculations agree strikingly well with the experiments. This is also evident from Fig. 2(e), where we show the correspondence between the experimental bias voltages and the theoretical energies. This gives the Fermi velocity $v_\mathrm{F}=(6.2\pm 0.1)\times 10^5$ m/s as the best-fit to both the KG equation and the TB calculations. The two theories yield slightly different values for the doping of the GQD, i.e. the intercept of the $y$-axis, due to the differences in the theoretical approaches.

\begin{figure}
\includegraphics{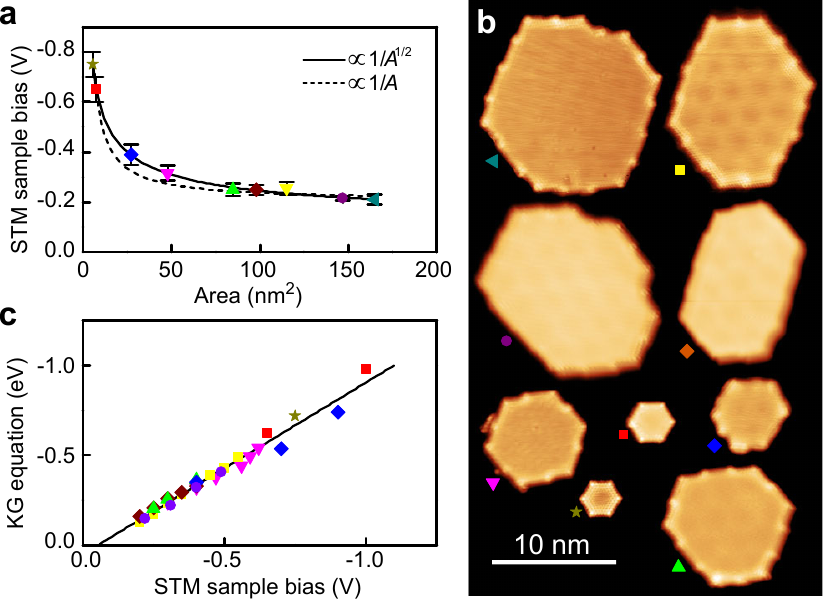}
\caption{(Color online) Electronic structure of GQDs as a function of their size. (a) STM sample bias corresponding to the $S$ state as a function of the area $A$ of the GQD. The solid and dashed lines are fits to $1/A^{1/2}$ and $1/A$ scaling, respectively. (b) Composition of the STM topographies of the GQDs used in panel (a) (with different scan parameters). (c) A plot of the bias voltages from the STM experiments ($x$-axis) and the energies that give comparable LDOS calculated from the Klein-Gordon equation using a single value for $v_\mathrm{F}=6.2\times10^5$ m/s ($y$-axis).}\label{Fig3}
\end{figure}
Do we see the peculiar nature of the charge carriers in graphene in these LDOS maps? In fact, the Schr\"odinger equation predicts wavefunctions with an identical spatial shape as the KG equation since both are second order differential equations; the corresponding eigenenergies are related as $E_\mathrm{S} = E_\mathrm{KG}^2/2mv_\mathrm{F}^2$. This also explains the different dispersion relations for free electrons, which are either parabolic (Schr\"odinger) or linear (Klein-Gordon). Moreover, the energy of the lowest (and the other) quantum confined state scales as $1/A^{1/2}$ ($A$ is the area of the QGD) in the case of the relativistic massless particles, instead of $1/A$ for the particles obeying the Schr\"odinger equation. We demonstrate in Fig. 3 that the charge carriers in our GQDs fulfil the conditions of $E\propto 1/A^{1/2}$ and have a linear dispersion. Fig. 3(a) shows the bias voltage corresponding to the lowest quantum confined energy level (determined by the peak position in $dI/dV_\mathrm{b}$ vs. $V_\mathrm{b}$ spectra acquired at the center of the GQD) on many different GQDs [topographies shown in Fig. 3(b)] as a function of the experimentally determined area. The solid line showing the expected $1/A^{1/2}$ scaling fits the data clearly better than the $1/A$ (dashed line) behavior.

In Fig. 3(c), we present the correspondence between experimental bias voltages ($x$-axis) and the theoretical energies calculated with the KG equation ($y$-axis) for many states on several GQDs. The one-to-one correspondence confirms that the experimental data is consistent with the linear dispersion of the Klein-Gordon equation. The corresponding Fermi velocity $v_\mathrm{F} = (6.2\pm 0.3)\times 10^5$ m/s is slightly smaller than the previous results on macroscopic graphene samples on Ir(111) obtained by ARPES ($6.5\times10^5$ to $9.2\times10^5$ m/s) \cite{Pletikosic2009,Rusponi2010,Starodub2011}. Possible reasons for this discrepancy are that our STM measurements probe the average Fermi velocity around the Dirac cone (in contrast to ARPES) and our experiments are carried out on GQDs instead of bulk graphene. Remarkably, $v_\mathrm{F}$ remains constant down to the smallest structures that we have measured. The intercept with the $y$-axis in Fig. 3(c) and the extrapolation to infinite GQD area in Fig. 3(a) indicate that GQDs on Ir(111) are n-doped by $\sim$ 0.1 eV.

In summary, we have presented low-temperature STM and STS experiments aimed at understanding the quantum confined energy levels and their spatially resolved wavefunctions in atomically well-defined graphene quantum dots. The measured resonances and corresponding LDOS maps correspond to a number of molecular orbitals close in energy, calculated by TB for the exact atomic geometry. The energy position and LDOS structure of these clustered states can also be calculated from the relativistic wave-equation for massless particles. Our results provide experimental verification of the physics relevant for graphene-based opto-electronics where wavefunction engineering via well-defined nanostructuring is likely to be a central issue. In addition, our experiments indicate that the intrinsic electronic states of graphene can be studied on weakly interacting metal substrates (e.g. Ir(111)). These systems can act as future test beds for studying the effects of chemical modifications or doping of graphene.

\begin{acknowledgments}
This research was supported by the Academy of Finland (Projects 117178, 136917, and the Centre of Excellence programme 2006-2011), FOM ("Control over Functional Nanoparticle Solids (FNS)"), the Finnish Academy of Science and Letters, and NWO (Chemical Sciences, Vidi-grant 700.56.423).

\end{acknowledgments}

\emph{Note added.}--During the review of this letter, we became aware of related experiments presented in Refs. \cite{Subramaniam2011,Phark2011}.

%\bibliographystyle{h-physrev3}
%\bibliographystyle{apsrev}
%\bibliography{refs}

\end{document}

\title{Supplementary Information: \\Quantum confined states in atomically well-defined graphene nanostructures}
\author{Sampsa K. H\"{a}m\"{a}l\"{a}inen }
\affiliation{Department of Applied Physics, Aalto University School of Science, 00076 Aalto, Finland}
\author{Zhixiang Sun}
\author{Mark P. Boneschanscher}
\affiliation{Condensed Matter and Interfaces, Debye Institute for Nanomaterials Science,
Utrecht University, PO Box 80000, 3508 TA Utrecht, the Netherlands}
\author{Andreas Uppstu}
\affiliation{Department of Applied Physics, Aalto University School of Science, 00076 Aalto, Finland}
\author{Mari Ij\"as}
\author{Ari Harju}
\affiliation{Department of Applied Physics, Aalto University School of Science, 00076 Aalto, Finland}
\affiliation{Helsinki Institute of Physics, Aalto University School of Science, FI-00076 Aalto, Finland}
\author{Dani\"{e}l Vanmaekelbergh}
\affiliation{Condensed Matter and Interfaces, Debye Institute for Nanomaterials Science,
Utrecht University, PO Box 80000, 3508 TA Utrecht, the Netherlands}
\author{Peter Liljeroth}
\affiliation{Department of Applied Physics, Aalto University School of Science, 00076 Aalto, Finland}
\affiliation{Condensed Matter and Interfaces, Debye Institute for Nanomaterials Science,
Utrecht University, PO Box 80000, 3508 TA Utrecht, the Netherlands}
\affiliation{Low Temperature Laboratory, Aalto University School of Science, PO Box 15100, 00076 Aalto, Finland}

\maketitle

\section{Orientation and edge structure of graphene quantum dots}
\begin{figure}[h!]
\includegraphics[width=.8\textwidth]{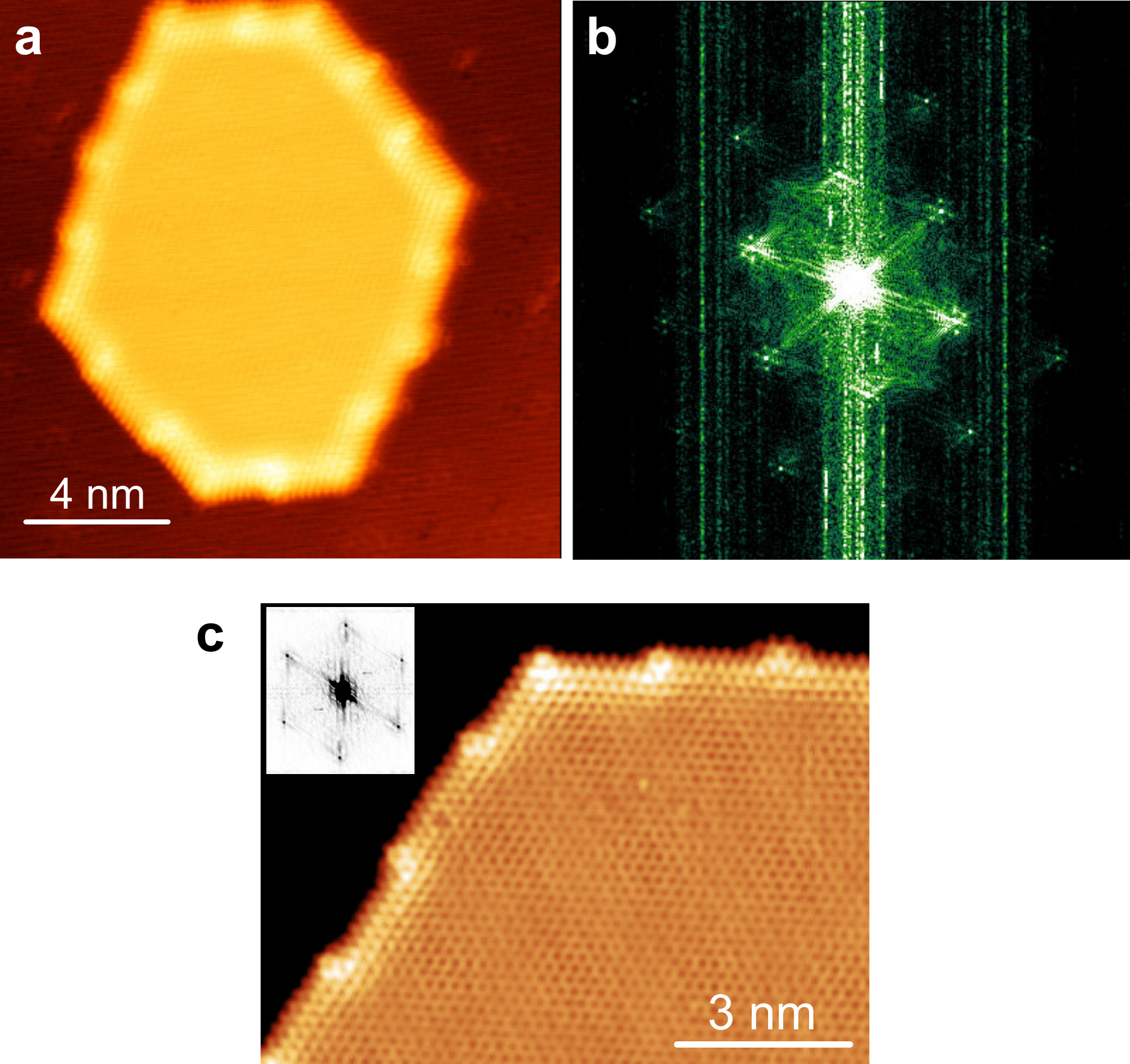}
\caption{STM images and their Fourier transforms showing GQD edge orientation and the alignment of the Ir and graphene lattices. (a) Atomically resolved STM image of the GQD presented in Fig. 2 of the main text (5 mV / 1nA). The contrast is enhanced by an adsorbed molecule on the tip, which also inverts the atomic contrast. (b) Square root of the modulus of the FFT of the panel (a). (c) Atomic resolution STM image (50 mV / 200 pA) of a larger graphene flake edge where the kinks of one atomic row are clearly visible. The inset shows the FFT spectrum of the image.}\label{structure}
\end{figure}

The lattices of all the observed graphene islands were roughly aligned with the underlying Ir lattice, which is the predominant growth phase of graphene on Ir \cite{Coraux2008}. This can be clearly seen in atomically resolved STM images and their Fourier transforms [Supplementary Figs. \ref{structure}(a) and \ref{structure}(b)]. The Fourier transforms show intensity maxima in a hexagonal pattern around the maxima produced by the graphene lattice. These spots arise from the lattice mismatch between graphene and iridium and are analogous to the diffraction pattern of graphene on Ir observed in low-energy electron diffraction (LEED) \cite{NDiaye2008}. In our case, these hexagons are aligned with the maxima from the graphene lattice indicating that the two lattices are aligned. The mismatch in lattice constants is also observed in the real space STM images as a moir\'e pattern on graphene islands with a period of around 2.5 nm \cite{NDiaye2008,Sun2011}.

The orientation of the GQD edges was determined from atomic resolution STM images and FFT spectra. All edges were terminated in the zigzag direction. This could be verified from FFT spectra of the GQDs, where the first order maxima were always oriented perpendicular to the graphene flake edges [Supplementary Fig. \ref{structure}(c)]. The edges always had 120 degree corners; no 60 degree corners were observed. Thus all GQDs had an overall hexagonal shape and no triangular QDs were observed.

\section{Edge states in small graphene QDs}
\begin{figure}[h!]
\includegraphics[width=.8\textwidth]{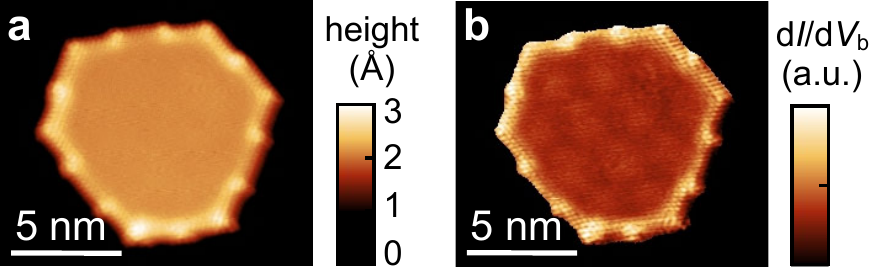}
\caption{STM measurements on edge states in small GQDs. (a) STM topography of an isolated GQD flake ($I = 100$ pA / $V_\mathrm{b} = 0.05$ V). (b) Simultaneously acquired $dI/dV_\mathrm{b}$ map under constant-current STM feedback.}\label{edge}
\end{figure}
Closer inspection of a small GQD [Supplementary Figs. \ref{edge}(a) and (b)] at a bias voltage close to zero bias shows that the edges are bright both in the actual STM topography image [Supplementary Fig. \ref{edge}(a)] and in the simultaneously recorded $dI/dV_\mathrm{b}$ image [Supplementary Fig. \ref{edge}(b)]. The enhanced conductance is due to edge states that are expected for zigzag or reconstructed zigzag edges \cite{CastroNeto2009,Koskinen2008}. We observe a higher LDOS at the corner and kink sites compared to the other edge sites, while atomistic tight-binding modeling predicts highest intensity at the middle of the edges with vanishing intensity in the corners of the GQD \cite{Heiskanen2008}. This disagreement is likely to be related to the remaining coupling of the GQDs with the Ir(111) substrate. Angle-resolved photoemission (ARPES) experiments have shown that close to the Fermi-level, there is an onset of a Ir(111) surface state that interacts with the graphene layer \cite{Pletikosic2009, Starodub2011}. At negative bias voltages, but before the onset of the delocalized states in the interior of the GQD, the experimental results on the edge-state LDOS agree with the predictions from tight-binding calculations. Bias-dependent imaging indicates that the edge states have a very flat dispersion, i.e. their spatial (exponential) decay away from the edges of the GQD is roughly energy independent with a decay constant of ca. 0.5 nm. Generally, we can observe the intrinsic properties of GQDs with little interference from the substrate in the bias region below -0.1 V.

\section{Comparison of the experiments with tight-binding calculations for a small GQD}
\begin{figure}[h!]
\includegraphics[width=.75\textwidth]{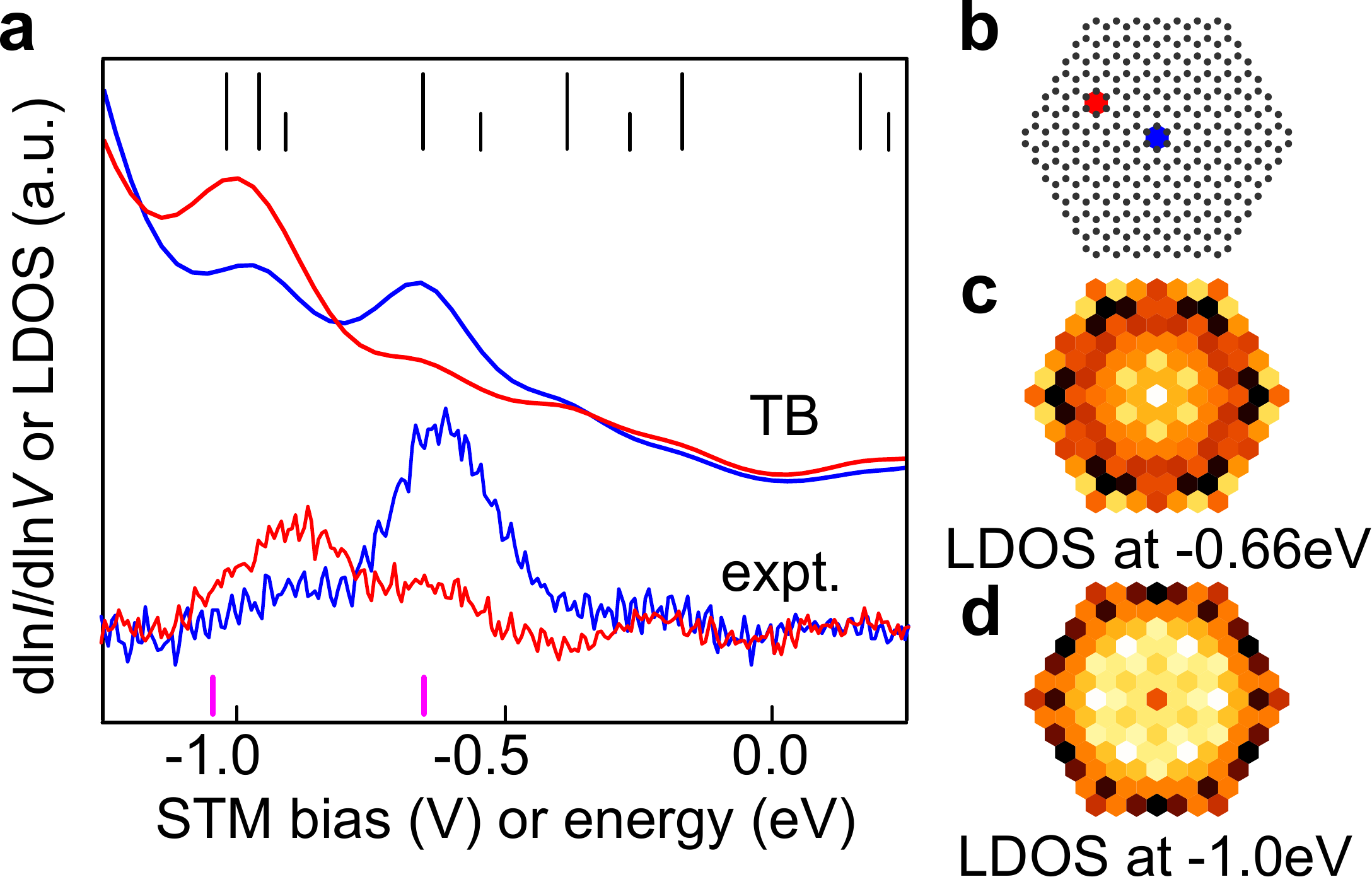}
\caption{(a) LDOS calculated by TB compared with experimental results reproduced from Fig. 1c of the main text. The TB energy levels are indicated by lines at the top of the graph (some levels are degenerate). The eigenenergies from the Klein-Gordon equation are indicated by magenta lines on the bottom of the graph. (b) Atomic model of the GQD indicating the positions where the LDOS is calculated in panel (a). (c) The calculated LDOS map based on TB at the energy corresponding to the 1S state. (d) calculated LDOS map based on TB at the energy corresponding to the 1P state.}\label{DOS}
\end{figure}

\section{Determination of the GQD shape and atomic structure}
\begin{figure}[h!]
\includegraphics[width=.8\textwidth]{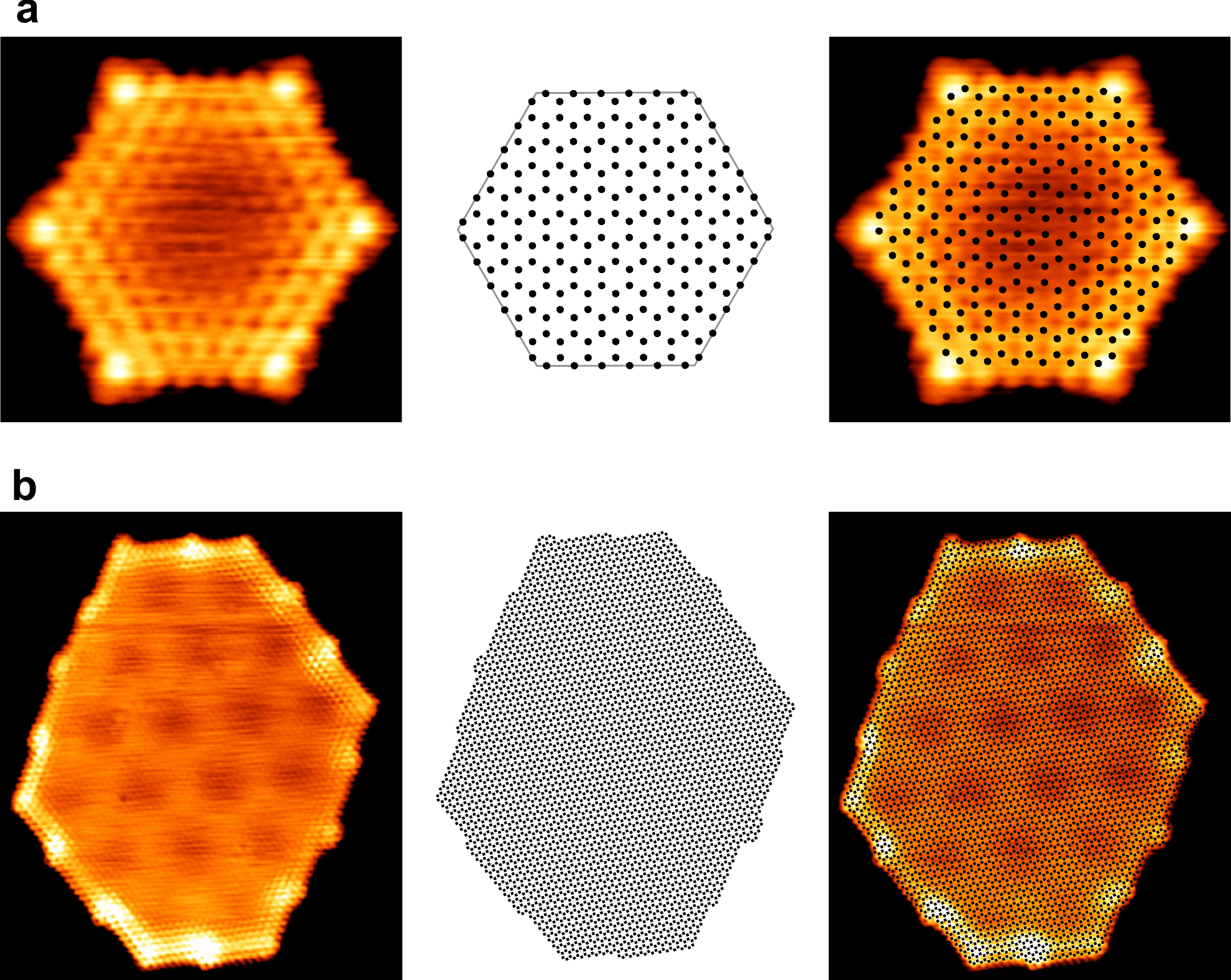}
\caption{Examples of the atomic models created for two different GQDs. (a) Atomic model created directly from the STM image for a GQD with 6 carbon rings per edge. The grey hexagon around the atomic model (middle column) represents the area used for the FEM calculation of the Klein-Gordon equation. (b) Example of a larger GQD where the atomic model has first been created from the coordinates of the edges and kinks and then refined to better match the actual atomic structure on the edges of the STM image.}\label{atomic}
\end{figure}

Atomic models of the GQDs were created for the TB calculations. For the smallest hexagonal GQDs, this procedure was straight-forward as the hexagons can be readily identified by counting the carbon rings along each edge of the GQD. An example of the smallest observed GQD with its corresponding atomic model is shown in Supplementary Fig. \ref{atomic}(a).

For the larger and more complex GQDs, a rough estimate of the structure was first created by measuring the coordinates of all the kinks and corners of the QGD from the STM image. An atomic model based on the coordinates was then created and compared to the atomically resolved image. As there might be an uncertainty in the STM piezo calibrations or some residual thermal drift, the model was then further refined by comparing the number of carbon rings in the image and model along the edges of the GQD. These models correspond to the actual GQD geometry with an uncertainty of 1 carbon atom row. The kinks and other features on the edge are nevertheless produced with atomic precision. An example of this is shown in Supplementary Fig. \ref{atomic}(b).

For solving the Klein-Gordon equation (which does not include atomic details), the shape and size were taken directly from the STM images for the larger QDs. The analysis of the GQD size and shape was done using the free SPM analysis software Gwyddion (http://gwyddion.net/). The kinks on the edges of the larger flakes were also taken into account. The symmetric smaller QDs were modeled as perfect hexagons with sharp corners, where the edges run along the outermost atoms of the atomic model [hexagon surrounding the atomic model in Supplementary Fig. \ref{atomic}(a)].

\section{$dI/dV_\mathrm{b}$ maps from Fig. 2b with the background}
\begin{figure}[h!]
\includegraphics[width=0.9\textwidth]{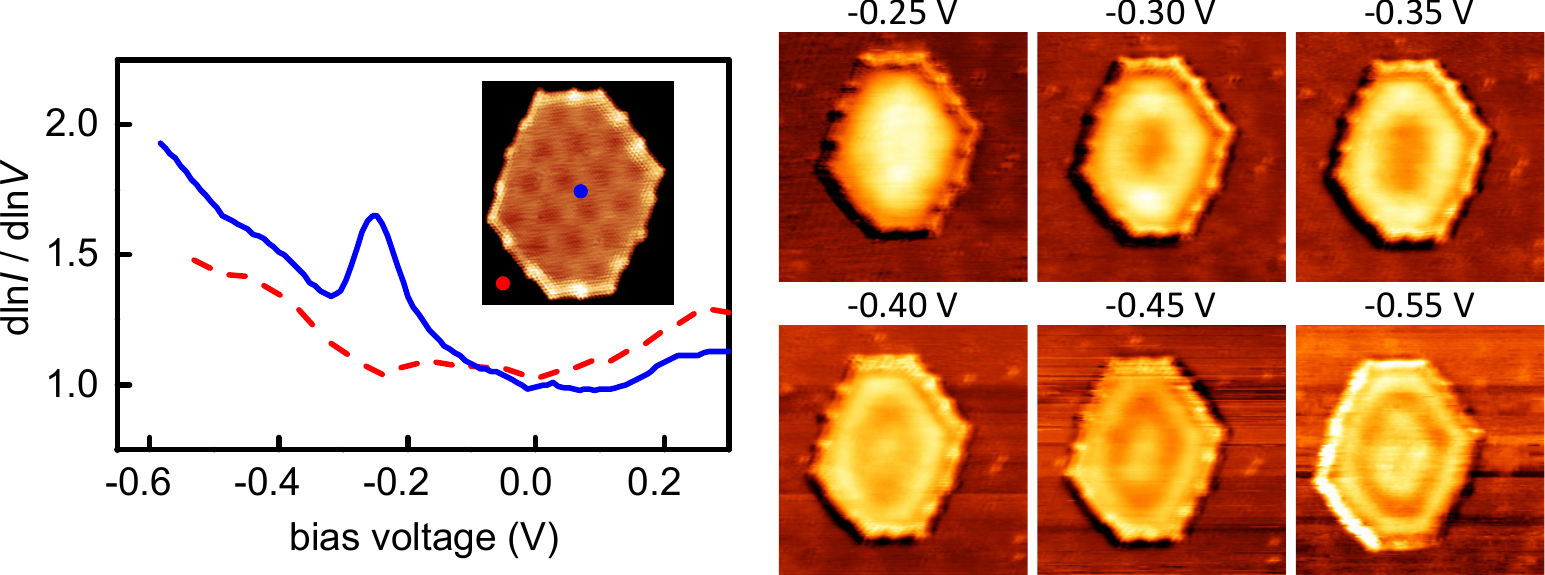}
\caption{Left: dln$I$/dln$V_\mathrm{b}$ spectra measured in the middle of the GQD shown in the inset (blue solid line) and on top of the Ir(111) substrate (red dashed line). This is the same GQD as shown in Fig. 2 of the main text. Right: Experimental $dI/dV_\mathrm{b}$ maps from Fig. 2b of the main text at the biases indicated in the figure showing the background, i.e. the surrounding Ir(111) substrate. Set-point current 1 nA.}\label{dIdVbackground}
\end{figure}

\newpage
\section{$dI/dV_\mathrm{b}$ maps measured at positive biases}
\begin{figure}[h!]
\includegraphics[width=.8\textwidth]{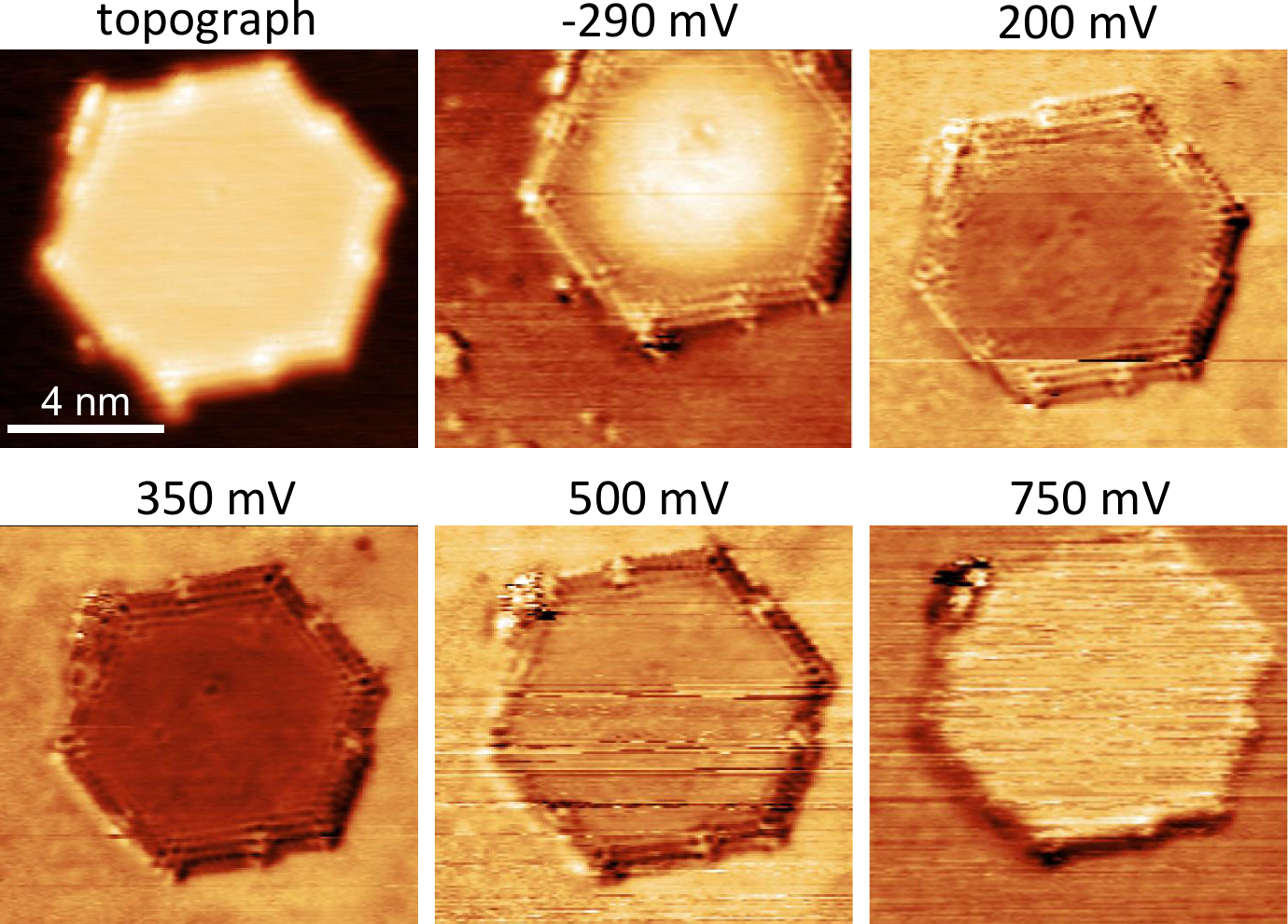}
\caption{$dI/dV_\mathrm{b}$ maps measured at positive bias showing no contrast over the graphene QD. Map corresponding to the energy of the 1S-state (bias voltage -0.29 V) is shown for comparison. Set-point current 0.5 nA. Topography of the GQD is shown on the top-right panel (bias voltage 0.1 V and set-point current 0.5 nA.}\label{positive}
\end{figure}

\section{Calculation of the LDOS from the Klein-Gordon equation}
The use of the Klein-Gordon equation to describe the quantum confined states in graphene QDs can be motivated starting from the Dirac equation \cite{CastroNeto2009}
\begin{equation}
-iv_\mathrm{F}\hbar\vec{\sigma}\cdot\vec{\nabla}\psi=E\psi
\end{equation}
where $\vec{\sigma}$ is the Pauli matrix and $\psi=\binom{\psi_a}{\psi_b}$ is wavefunction for the two sublattices a and b. Writing this out in component form gives
\begin{equation}
\begin{array}{rl}
-iv_\mathrm{F}\hbar(\partial_x - i\partial_y)\psi_b=E\psi_a\\
-iv_\mathrm{F}\hbar(\partial_x + i\partial_y)\psi_a=E\psi_b
\end{array}
\end{equation}
Separating the components gives the Klein-Gordon equation for both sublattices
\begin{equation}
\begin{array}{rl}
-v_\mathrm{F}^2\hbar^2\nabla^2\psi_a=E^2\psi_a\\
-v_\mathrm{F}^2\hbar^2\nabla^2\psi_b=E^2\psi_b\\
\end{array}
\end{equation}
For an infinite graphene flake (i.e. ignoring the effect of boundary conditions), all the solutions of the Dirac equation are also solutions of the Klein-Gordon equation.

The Klein-Gordon equation was solved numerically by the finite element method (FEM) using the commercial software Comsol Multiphysics (v. 3.5). Eigenstates of the Klein-Gordon equation were calculated up to 2 eV (using $v_\mathrm{F} = 6.2\times10^5$ m/s). LDOS maps for a given energy $E$ were created by weighing each eigenstate $\psi_i$ by the energy difference of the eigenvalue $E_i$ and $E$ using a Gaussian distribution
\begin{equation}
LDOS(E,x,y)=\sum_i \left [e^{-\frac{(E-E_i)^2}{2\sigma^2}}\psi(x,y)^2_i \right]
\label{spread}
\end{equation}
We checked the effect of the width $\sigma$ on the calculated LDOS maps: 60 meV was found to give the best match with the experimental results and was thus used for all of the calculations.

\section{Tight-binding calculations}
We used a single-electron tight-binding (TB) model, suitable for describing the $\pi$-electrons of graphene. Hoppings up to third-nearest neighbors were included, with the original parameters being $t_1 = -2.7$ eV, $t_2 = -0.20$ eV and $t_3 = -0.18$ eV for first, second and third-nearest neighbors, respectively \cite{Hancock2010}. In order to match the experimental bias voltages, the parameters were scaled to $t_1 = -2.26$ eV, $t_2 = -0.168$ eV and $t_3 = -0.151$ eV, keeping their ratios constant. In 2D graphene, these parameters correspond to a Fermi velocity of about $6.3\times10^5$ m/s. The LDOS was computed with a 60 meV Lorentzian broadening of the spectrum. In the figures, the LDOS from the TB model has been averaged over the sites in each hexagon of the graphene lattice.

\section{Visual comparison of measured and calculated maps}
\begin{figure}
\includegraphics[width=.8\textwidth]{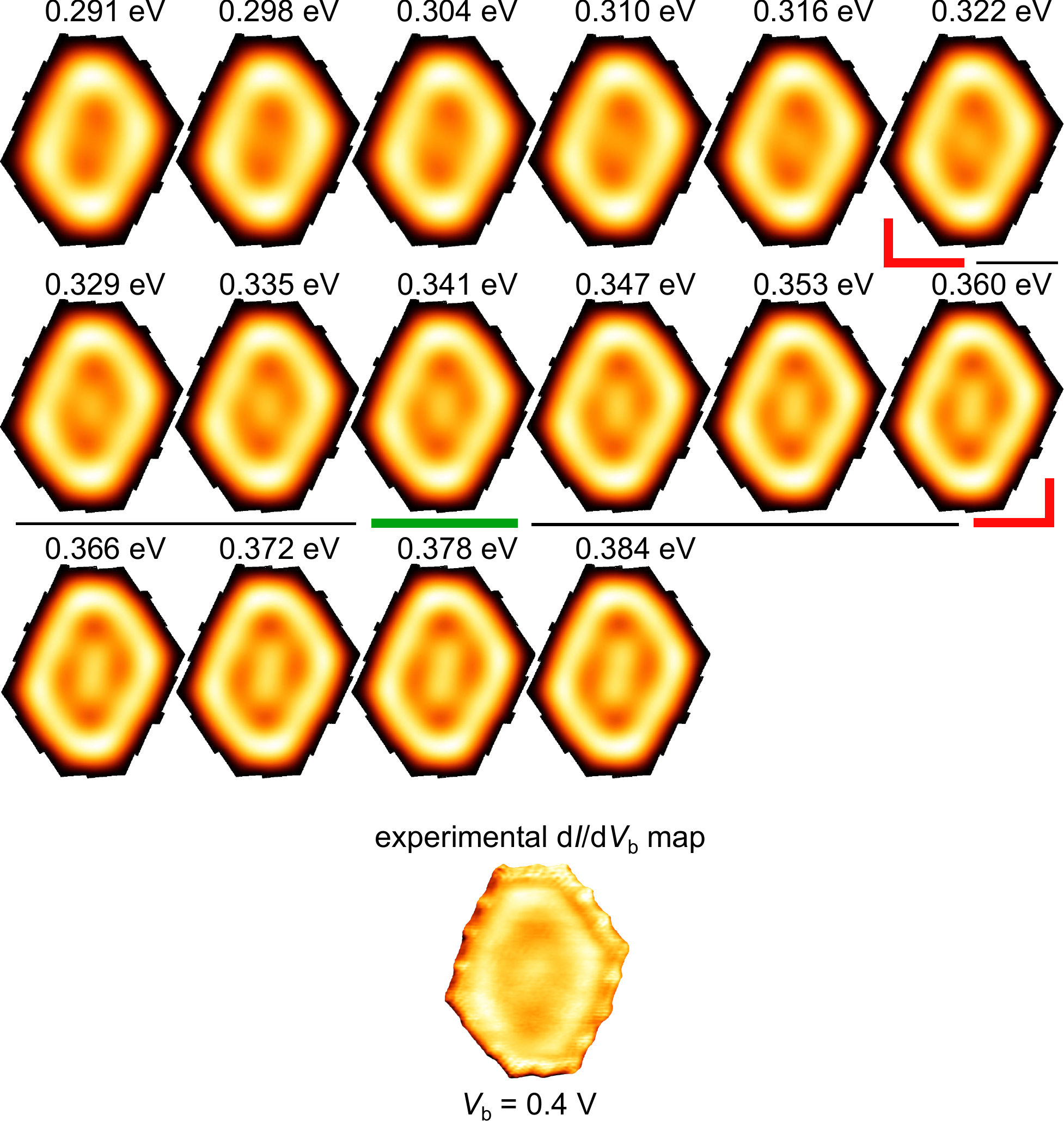}
\caption{LDOS maps calculated based on the KG equation. Map at the energy of 0.341 eV (green underline) is taken as the best match to the measured $dI/dV_\mathrm{b}$  map. The red markers denote the limits of error for the visual match.}\label{visual}
\end{figure}
To find the correspondence between the measured and calculated states the LDOS maps were visually compared. This was done by calculating several LDOS maps in relatively small energy steps (6 meV) and comparing these to each of the measured $dI/dV_\mathrm{b}$ maps. Since in most cases the calculated states change fairly little between such small energy steps, a range of possible matches was picked for each measured $dI/dV_\mathrm{b}$ map. The energy in the middle of the range was used as the best match. The width of the range of the corresponding experimental and calculated LDOS maps was typically 50 meV. An example of this procedure is shown in Supplementary Fig. \ref{visual}, where the best match is indicated by the green bar.

Subsequently, the Fermi velocity has been scaled such that the slope of the plot of the theoretical energies vs. the experimental bias voltage is equal to unity. This gives the best-fit Fermi velocity. In the case of the TB calculations (which has three hopping parameters), we scale all the hopping parameters such that their ratios remain constant. The intercept of the plot of the theoretical energies vs. the experimental bias voltages gives the doping of our graphene in the bulk limit. It is not a freely adjustable parameter but rather a result of the comparison between theory and experiment.

%\bibliographystyle{h-physrev3}
%\bibliographystyle{apsrev}
%\bibliography{refs}